\documentstyle[./bo99,epsfig]{article}

\def\etal{{\it et al.\/}}

\def\ie{{\it i.e.\/}}

\title{ What have learned about Gamma Ray Bursts from afterglows?}
\author{ Mario Vietri}
\affil{ Universit\`a di Roma III}

\begin{document}

\maketitle

\begin{abstract}
The discovery of GRBs' afterglows has allowed us to establish several facts:
their distance and energy scales, the fact that they are due to
explosions, that the explosions are relativistic, and that the afterglow
emission mechanism is synchrotron radiation. On the other hand, recent
data have shown that the fireball model is wrong when it comes to the
emission mechanism of the true burst (which is unlikely to be synchrotron
again) and that shocks are not external. Besides these relatively tame points,
I will also discuss the less well established physics of the energy 
deposition mechanism, as well as the possible burst progenitors. 
\end{abstract}

\section{Introduction}

Gamma ray bursts (GRBs) were discovered in 1969 (Klebesadel, Strong and Olson
1973) by American satellites of the {\it Vela} class aimed at verifying 
Russian compliance with the nuclear atmospheric test ban treaty. Though
the discovery was made in 1969, the paper appeared only four years later 
because the authors had lingering 
doubts about the reality of the effects they had discovered. Since then,
several thousands of bursts have been observed by a more than a dozen 
different satellites, but it is remarkable that the basic burst features
outlined in the abstract of the 1969 paper (photons in the range 
$0.2-1.5\; MeV$, durations of $0.1-30\; s$, fluences in the range 
$10^{-5}-2\times 10^{-4}\; ergs\; cm^{-2}$) have remained substantially 
unchanged. 

Current evidence (Fishman and Meegan 1995) has highlighted a wide ($0.01-100\; 
s$) duration distribution, with hints of a bimodality which is
claimed to correlate (at the $2.5\sigma$ level) with spectral properties. 
All bursts' spectra observed so far are strictly non--thermal, and there has 
never been any confirmation by BATSE of a supposed thermal component (nor of 
cyclotron lines or precursors, for this matter) claimed in previous reports.
A remarkable feature reported by BATSE is the bewildering diversity of
light curves, ranging from impulsive ones (a spike followed by
a slower decay, nicknamed FREDs for Fast Rise-Exponential Decay), to 
smooth ones, to long ones with amazingly sharp fluctuations, including 
even some with a strongly periodic appearance (two such examples are the 
`hand' and the `comb', so nicknamed from the number of high--Q, regularly
repeating sharp spikes). 

The most exceptional result from BATSE, though, was the sky distribution 
of the bursts (Fig.1). It was obvious from it that the bursts {\it had}
to be extragalactic, as already discussed by theorists (Usov and Chibisov 1975, 
Paczy\`nski 1986). 
\begin{figure}
\centerline{\psfig{file=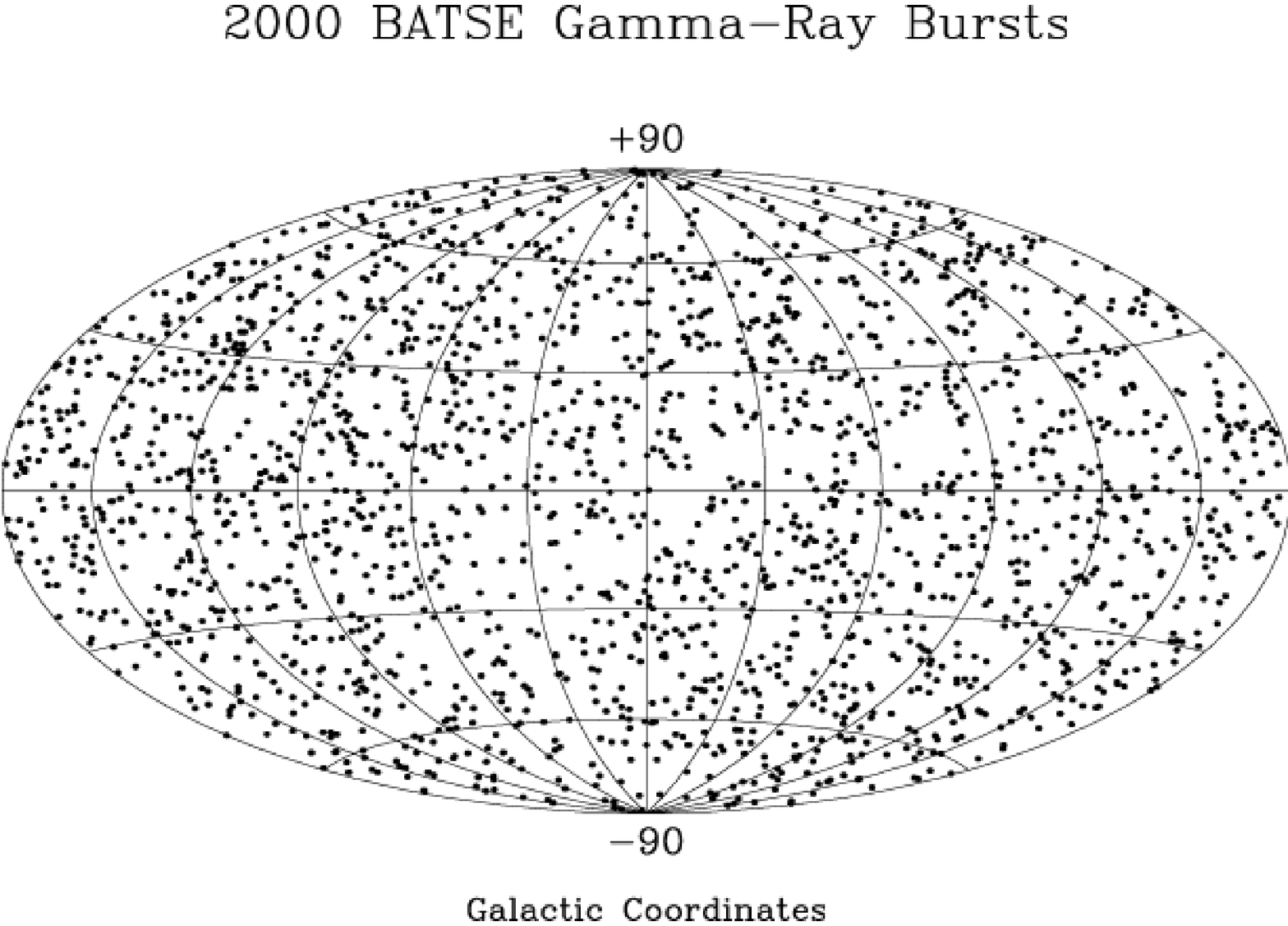, width=10cm}}
\caption[]{Burst distribution on the plane of the sky,}
\end{figure}

\section{Afterglows}

The next major step was triggered by BeppoSAX: in the summer of 1996, L. Piro
and his coworkers located in archival data of the satellite the soft 
X--ray counterpart of a GRB (GRB 960720). They immediately conceived the 
idea of implementing a procedure to follow the next burst in real time, by 
re-orienting the whole satellite, after the initial detection by the
Wide Field Cameras, so that the more sensitive Narrow Field Instruments 
could pinpoint the burst location to within $45$ arcsecs, a feat never 
achieved in such short times, and by a single satellite. After an initial 
snafu (GRB 970111), the gigantic effort paid off with the discovery of 
the X--ray afterglow of
GRB 970228 (Costa \etal, 1997), immediately followed by the discovery of its
fading optical counterpart (van Paradijs \etal, 1997), obtained through a
search inside the WFC error box, in perfect agreement with theoretical 
predictions (Vietri 1997a, M\`esz\`aros and Rees 1997)\footnote{There is an 
interesting lesson to be drawn from this: in case one should wonder why 
a soft X--ray telescope was not placed onboard Compton to track GRBs,
it was because of rivalries between different NASA subsections, the 
$X$--ray and the $\gamma$--ray divisions.}. 

After the detection of the optical counterpart, the door was open to find
the bursts' redshifts: Table I summarizes the status of our current 
knowledge (september 1999); bursts' luminosities are for isotropic
sources. 
\begin{table}[t]
\begin{center}
\begin{tabular}{|c|c|c|} \hline
GRB & z & $E_{iso}$ \\ \hline
970228 & 0.695 & $5\times 10^{51}\; erg$ \\ \hline
970508 & 0.835 & $2\times 10^{51}\; erg$ \\ \hline
971214 & 3.4 & $3\times 10^{53}\; erg$ \\ \hline
980703 & 0.93 & $3\times 10^{53}\; erg$ \\ \hline
990123 & 1.7 & $4\times 10^{54}\; erg$ \\ \hline
990510 & 1.6 & $2\times 10^{53}\; erg$ \\ \hline
990712 & 0.43 &  \\ \hline
\end{tabular}
\end{center}
\end{table}
Two comments are in order. 
First, the bursts have {\it prima facie} a redshift distribution not 
unlike that of AGNs and of the 
Star Formation Rate (SFR). The initial hope that they might trace an even
more distant and elusive Pop III, triggered by the fact that the second
redshift detected was also the largest so far (GRB 971214, $z=3.4$), has now
vanished. Second, in order to place the energy release of GRB 990123 in 
context, one should notice that $4\times 10^{54}\; ergs$ is the energy
obtained by converting the rest--mass of two solar masses, or, alternatively,
the energy emitted by the whole Universe out to $z\approx 1$ within
the burst duration. So, a single (perhaps double) star outshines the whole
Universe.

Besides the distance and energy scales, the major impact of the discovery 
of afterglows has been the establishment of some key features of the 
fireball model (Rees and M\`esz\`aros 1992):
\begin{enumerate}
\item bursts are due to explosions, as evidenced by their power--law behaviour;
\item the explosions are relativistic, as proved by the disappearence
of radio flares;
\item the burst emission is due to synchrotron emission, as shown by the
afterglow spectrum, and its optical polarization.
\end{enumerate}
I will illustrate these points in the following, but, lest we become too proud, 
we should also remember that the fireball model has met some failures. The 
original version of the model (M\`esz\`aros and Rees 1993) advocated the
dissipation of the explosion energy at external shocks (\ie, those with the
interstellar medium). Sari and Piran (1997), following a point originally
made by Ruderman (1975) showed that these shocks smooth out millisecond
timescale variability, which can only be maintained by the internal shocks 
proposed by Paczy\`nski and Xu (1994). Also, the fireball model originally 
ascribed even the emission from the burst proper (as opposed to the 
afterglow) to optically thin synchrotron processes; I will discuss in the
section {\it Embarrassments} why this is exceedingly unlikely. Furthermore,
even the last tenet of mid--90s common wisdom, \ie, that bursts are due to
neutron binary mergers, does not look too promising at the moment (since some
bursts seem to be located inside star forming regions, incompatible with 
the long spiral--in time), though of course it is by no means ruled out yet. 

\subsection{The fireball model}

Here, one may assume that an unknown agent deposits $10^{51}-10^{54}\; ergs$
inside a small volume of linear dimension $\approx 10^6-10^7\;cm$. The resulting
typical energy density corresponds to a temperature of a few $MeV$s, so that 
electrons and positrons cannot be bound by any known gravitational 
field. In these conditions, optical depths for all known processes exceed
$10^{10}$. The fluid expands because of its purely thermal pressure,
converting internal into bulk kinetic energy. Parametrizing the baryon 
component mass as $M_b \equiv E/\eta c^2$, it can be shown that, for 
$1 \leq \eta \leq 3\times 10^5$ (M\`esz\`aros, Laguna, Rees 1993) the fluid 
achieves quickly (the fluid Lorenz factor increases as $\gamma \propto r$) a 
coasting Lorenz factor of $\gamma \approx \eta$. 

The requisite asymptotic Lorenz factor is dictated by observations: photons
up to $\epsilon_{ex} \approx 18\; GeV$ have been observed by EGRET from bursts 
(Fishman and Meegan 1995). For these photons to evade collisions with other 
photons, and thus electron/positron pair production, it is 
necessary that, in the reference frame in which a typical burst photon (with 
$\epsilon \approx 1\; MeV$) and the exceptional photon are emitted, they appear 
as below pair production threshold: thus we must have $\epsilon' \epsilon_{ex}'
\leq 2 m_e c^2$. Since $\epsilon' \approx \epsilon/\gamma$, and similarly
for the other photons, we find (Baring 1993)
\begin{equation}
\gamma \approx 300 \left(\frac{\epsilon}{1\; MeV}\frac{\epsilon_{ex}}{10\; GeV}
\right)^{1/2}\;.
\end{equation}
From what we said above, we thus require a maximum baryon contamination, in an
explosion of energy $E$, of $M_b \la 10^{-6} M_\odot (E/10^{51}\; erg) 
(300/\eta)$. 

The energy release is now assumed to be in the form of an inhomogeneous
wind, with parts having a Lorenz factor larger than parts emitted 
previously. This leads to shell collisions (the internal shock model) at 
radii $r_{sh}$ which allow a time--scale variablity $\delta\!t \approx
 r_{sh}/2\gamma^2 c$; for $\delta\!t = 1\;ms$, $r_{sh} \approx 10^{13}\; cm$,
which fixes the internal shock radii. Particle acceleration at these 
internal shocks and ensuing non--thermal emission is thought to lead to the 
formation of the burst proper. At larger radii, a shock with the
surrounding ISM forms, and shell deceleration begins at a radius $r_{ag} = 
(3E/4\pi n m_p c^2\gamma^2)^{1/3} \approx 10^{17}\; cm$ for a $n = 1\; cm^{-3}$ 
particle density typical of galactic disks. It is thought that the afterglow 
begins when the shell begins the slowdown, as this drives a marginally 
relativistic shock into the ejecta, thusly extracting a further fraction of 
their bulk kinetic energy. 

\subsection{Why explosions}

The success of the fireball model lies in this, that it decouples the problem
of the energy injection mechanism from the following evolution, which is, 
furthermore, an essentially hydrodynamical problem. It can be shown, in fact 
(Waxman 1997) that the evolution of the external shock is adiabatic, 
that the shock Lorenz factor decreases as $\gamma \propto r^{-3/2}$ 
because of the inertia of the swept--up matter, and thus $r$ scales with
observer's time as $t = r/\gamma^2 c \rightarrow \gamma \propto t^{-3/8}$
(for a radiative solution $\gamma \propto r^{-3/7}$, Vietri 1997b). If 
afterglow emission is due to optically thin synchrotron in a magnetic field in 
near--equipartition with post--shock energy density, it can be shown that $B 
\propto \gamma$, that the typical synchrotron frequency at the spectral peak 
$\nu_m \propto \gamma B \gamma_e^2 \propto \gamma^4$ (where $\gamma_e \propto
\gamma$ is the lowest post--shock electron Lorenz factor), and that $F(\nu_m) 
\propto t^{-3\beta/2}$, where $\beta$ is the afterglow spectral slope. As it
can be seen, these expectations are based exclusively upon the hydrodynamical
evolution (and the synchrotron spectrum), and are thus reasonably robust.

We thus expect power--law time decays, a characteristic of strong explosions
(see the Sedov analogy!), with time-- and spectral--indices closely related. 
This is what is observed everywhere, from the X--ray through the optical to the 
radio, (see Piro and Fruchter, this volume), the few exceptions being discussed 
later on. In fact, the equality of the time--decay
index of the X--ray and optical data in afterglows of individual sources has
been taken as the key element to show that emission in the different bands is
due to the same source. Time indices  in the X--ray are in the range 
$0.7-2.2$ (Frontera \etal, in preparation). 

\subsection{Why synchrotron spectrum in the afterglow}

After having established that bursts are due to explosions, we happily learn
that afterglows emit through synchrotron processes. In fig. 2 (Galama
\etal, 1998), we show the superposition of theoretical expectations for
an optically thin synchrotron spectrum (including the cooling break at 
$\nu \approx 10^{14}\; Hz$) with observations for GRB 970508. 
\begin{figure}
\centerline{\psfig{file=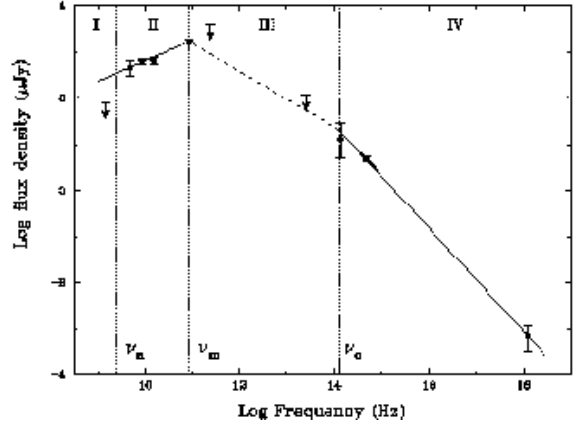, width=10cm}}
\caption[]{Simultaneous spectrum of the afterglow of GRB 970508, from
Galama \etal, 1998.}
\end{figure}
The remarkable
agreement is even more exciting as we remark that observations are not truly
simultaneous, but are scaled back to the same time by means of the 
theoretically expected laws for time--decay, thus simultaneoulsy testing the
correctness of our hydro. Another piece of evidence comes from the discovery
of polarization in the optical afterglow of GRB 990510 (Fig. 3, Covino \etal, 
1999, Wijers \etal, 1999). 
\begin{figure}
\centerline{\psfig{file=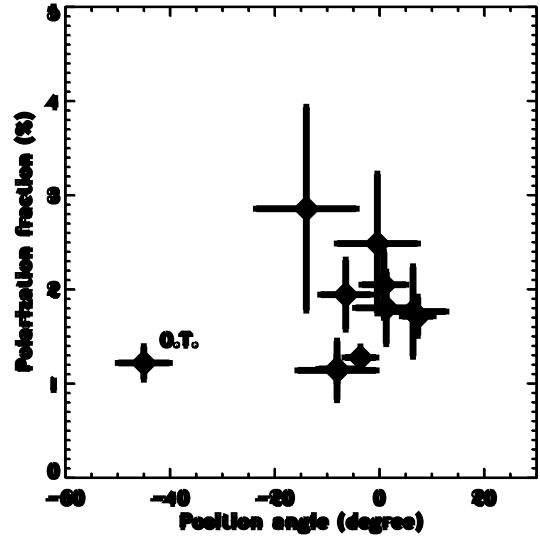, width=7cm}}
\caption[]{Polarization amplitude and position angle for optical afterglow
of GRB 990510, from Covino \etal, 1999}
\end{figure}
This polarization may appear small ($\approx 2\%$), but 
it is surely not due to Galactic effects: stars in the same field show a 
comparable degree of polarization, but along an axis different by about 
$50^\circ$. Also, polarization in the source galaxy is unlikely, because of
a very stringent upper limit on the reddening due to this galaxy (Covino \etal,
1999). The only remaining question mark is emission from an anisotropic source,
but this would require a disk of $10^{18}\; cm$ to survive the intense $\gamma$
ray (and X, and UV) flash: though not excluded, it does not look likely. 

\subsection{Why relativistic expansion}

Radio observations of the first burst observed so far (GRB 970508, Frail
\etal, 1997) showed puzzling fluctuations by about a factor of $2$ in the flux,
over a time--scale of days, disappearing after about $30$ days from the burst
(Fig. 4). 
This extreme, and unique behaviour, was explained by Goodman (1997), who showed
that it is due to interference of rays travelling along different paths through
the ISM, and randomly deflected by the spatially varying refractive index of
the turbulent ISM. The wonderful upshot of this otherwise marginal phenomenon, 
is that these effects cease whenever the source expands beyond a radius
\begin{equation}
R = 10^{17}\; cm \frac{\nu_{10}^{6/5}}{d_{sc,kpc} h_{75}} \left(
\frac{SM}{10^{-2.5} m^{-20/3}\; kpc}\right)^{-3/5}\;,
\end{equation}
where $\nu_{10}$ is the radio observing frequency in units of $10^{10} \; Hz$,
$d_{sc,kpc}$ is the distance of the ISM from us (assumed to be a uniform 
scattering screen), and $SM$ is the Galactic scattering measure, scaled to a
typical Galactic value. 
\begin{figure}
\centerline{\psfig{file=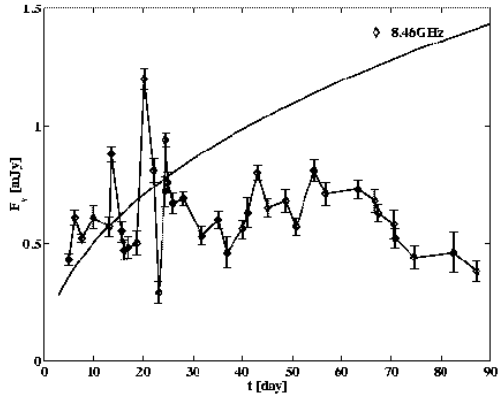, width=7cm}}
\caption[]{VLA observatons at $8.46\; GHz$ of the afterglow of GRB 970508,
from Waxman,Frail and Kulkarni 1998.}
\end{figure}
The existence of interference effects is made 
more convincing by the amplitude of the average increase (a factor of 2, as 
observed), the correctness in the prediction of the time--interval between
different peaks, and of the decorrelation bandwidth. Since flares 
disappear after about $30$ days, it means that the average speed of the
radio source is $R/30 {\mbox days} = 3\times 10^{10}\; cm\; s^{-1}$. So we 
see directly that GRB 970508 expanded at an average speed of $c$ over a whole
month, giving us a direct observational proof that the source is highly
relativistic. This proof is completely equivalent to superluminal motions in
blazars, and is the strongest evidence in favor of the fireball model.

\subsection{GRB 970508: our best case}

The afterglow of GRB 970508 is our best case so far: it is in fact a burst
for which not only do we know the redshift, but also a radio source that 
has been monitored for more than $400$ days after the explosion (Frail, 
Waxman and Kulkarni 2000). Through these observations we can see the transition 
to a sub--relativistic regime at $t \approx 100 \; d$, measure the total 
energetics of the following Sedov phase (unencumbered by relativistic effects!)
$E_{New} = 5\times 10^{50}\; ergs$, determine two elusive parameters, 
$\epsilon_{eq} 
= 0.5$ and $\epsilon_B= 0.5$ (the efficiencies with which energy is 
transfered to post--shock electrons by protons, and with which an 
equipartition field is built up), and the density of the surrounding medium
$n \approx 0.4\; cm^{-3}$. All of these values look reasonable (perhaps
$\epsilon_{eq}$ and $\epsilon_B$ exceed our expectations by a factor of 10,
a fact that could be remedied by introducing a slight density gradient which 
would keep the shock more efficient), so that our confidence in the 
external--shock--in--the--ISM model is boosted. 

Another precious consequence
of these late--time observations is that they yield information on 
beaming and energetics. In fact, GRB 970508 appeared to have a 
kinetic energy of $E_{rel} = 5\times 10^{51}\; erg$ when in the relativistic
phase, a measurement which can be reconciled with $E_{New}$ (remember that 
the expansion is adiabatic, so that we must have $E_{New} = E_{rel}$!) 
only if the unknown beaming angle, assumed $= 4\pi$ in deriving $E_{rel}$,
is smaller than $4\pi$ by the factor $E_{New}/E_{rel}$; we thus have the only
measurement of $\delta\!\Omega/4\pi = 0.1$, so far. This already rules out
all classes of models requiring unplausibly large amounts of beaming, $10^{-8}$
or even beyond. Hopefully, more 
such measurements will come in the future, since this observationally heavy
method is subject to many fewer uncertainties than the competing method of 
trying to locate breaks in the time--decay of afterglows. Also, the 
radiative efficiency of the burst can be estimated: correcting the 
observed burst energy release $E_{GRB} = 2\times 10^{51}\; erg$ for the same 
beaming factor, the radiative efficiency is $E_{GRB} 
\delta\!\Omega/4\pi /(E_{New}+E_{rel}\delta\!\Omega/4\pi) = 0.3$, again a
unique determination. Notice however that this figure is subject to a 
systematic uncertainty: we do not know whether the beaming fraction is the same
for the burst proper and for the afterglow. 

\section{Embarrassments}

Something is rotten in the fireball kingdom as well, namely, departures 
from pure power--law behaviours, and the spectra of the bursts proper.

\subsection{Unpowerlawness}

Departures from power--laws are expected when one considers the extremely
idealized character of the solutions discussed so far: perfect spherical
symmetry, uniform surrounding medium, smooth wind from the explosion, 
$\epsilon_{eq}$ and $\epsilon_B$ constant in space and time. The tricky
point here is to disentangle these distinct factors. In GRB 970508 and
GRB 970828 (Piro \etal, 1999, Yoshida \etal, 1999) a major departure was
observed in the X--ray emission, within a couple of days from the burst; they
constitute the single, largest violations observed so far, in terms of number of
photons. It is remarkable that spectral variations were simultaneously 
observed, and that both bursts showed traces (at the $2.7\sigma$ significance
level) of an iron emission line. The similarity of the bursts' behaviour 
argues in favor of the reality of these spectral features, which have been
interpreted as thermal emission from a surrounding stellar--size leftover, 
pre--expelled by the burst's progenitor (Lazzati \etal, 1999, Vietri \etal,
1999). Clearly, these departures hold major pieces of information on the 
bursts' surroundings, and the nature of bursts' progenitors. 

It has been argued (Rhoads 1997) that, whenever the afterglow shell 
decelerates to below $\gamma \approx 1/\theta$, where $\theta$ is the beam
semi--opening angle, emission should decrease because of the lack of emitting 
surface, compared to an isotropic source. But, in view of the existence of 
clear environmental effects (GRB 970508 and GRB 970828), it appears premature 
to put much stock in the interpretation of time--power--law breaks as due to 
beaming effects. And equally, it appears to this 
reviewer that the same comment applies to the interpretation of a resurgence
of flux as due to the appearance of a SN remnant behind the shell. The major
uncertainty here is the non--univoqueness of the interpretation: Waxman and 
Draine (2000) have shown that effects due to dust can mimic the same 
phenomenon. 

\subsection{Bursts' spectra}

A clear prediction of the emission of optically thin synchrotron is that 
the low--photon--energy spectra should scale like $d\!N_\nu/d\!\nu \propto
\nu^{\alpha}$, with $\alpha = -3/2$, since the emission is in the fast cooling
regime. Within thin synchrotron, there is no way to obtain $\alpha > -3/2$. 
This early--recognized requirement (Katz 1994) is so inescapable that it
has been dubbed the `line of death'. Observations are notoriously discordant
with this prediction. Preece \etal (1999) have shown that, for more than 1000
bursts, $\alpha$ is distributed like a bell between $-2$ and $0$, with mean
$\bar{\alpha} \approx -1$. The tail of this distribution also contains a few
tens of objects with $\alpha \approx +1$. An example of these can be found in
Frontera \etal, 1999 (GRB 970111), which is instructive since BeppoSAX 
has better coverage of the critical, low--photon--energy region. In particular,
BATSE seems to loose sensitivity below $\approx 30\; keV$, but this is still
not enough to explain away the discrepancy with the theory. Also, Preece
\etal, 1999, showed that the time--integrated spectral energy distribution
has a peak at a photon energy $\epsilon_{pk} \approx 200\; keV$, and that 
$\epsilon_{pk}$ has a very small variance from burst to burst. Again, this does 
not seem dependent upon BATSE's lack of sensistivity above $700\; keV$, and 
again this has no explanation within the classic fireball model. 

Any theorist who worked on blazars will say that the root of the disagreement 
is the neglect of Inverse Compton processes, but the trick here is not to
identify the culprit, on which everyone agrees, but to devise a fireball 
model that smoothly incorporates it. One should remember that the details of
the fireball evolution are {\it generic}, \ie, they do not depend upon any
detailed property of the source, so that things like the radius at which the 
fireball becomes optically thin (to pairs or baryonic electrons), the radius at 
which acceleration ends, the equipartition magnetic field, and so on, are all 
reliably and inescapably fixed by the outflow's global properties. A step 
toward the solution has been made by Ghisellini and Celotti (1999) who remarked
that at least some bursts have compactness parameters $l = 10 (L/10^{53}\;
erg\; s^{-1}) (300/\gamma)^5\gg 1$. Under these conditions, a pair plasma
will form, nearly thermalized at $kT \approx m_e c^2$, and with Thomson optical 
depth $\tau_T \approx 10$. The modifications which this plasma will bring to 
the burst's spectrum are currently unknown, but it may be remarked that this 
configuration will be optically thick to both high--energy synchrotron photons
due to non--thermal electrons accelerated at the internal shocks, and to 
low--energy cyclotron photons emitted by the thermal plasma, but it will be 
optically thin in the intermediate region reached by cyclotron photons 
upscattered via IC processes off non--thermal electrons. A model along
this line (\ie, upscattering of cyclotron photons by highly relativistic
electrons) is in preparation (Vietri 2000a), but it remains to be seen whether
it (like any other model, of course) can simultaneously explain the spectral
shape and the narrow range of the spectral distribution peak energy 
$\epsilon_{pk}$.

\section{On the central engine}

As remarked several times already, the fireball evolution is independent of
the source nature. The only exisiting constraint is the maximum amount of 
baryon contamination, which is
\begin{equation}
M_b = \frac{E}{\eta c^2} = 10^{-6} M_\odot \frac{E}{10^{51}\; erg}
\frac{300}{\gamma}\;.
\end{equation}
This is a remarkably small value: since the inferred luminosities exceed the
Eddington luminosity by 13 orders of magnitude, they clearly have all it takes
to disrupt a whole star, no matter how compact. Yet, the energy deposition must 
somehow occur outside the main mass, lest the explosion be slowed down to 
less relativistic, or even possibly Newtonian speeds. In order to satisfy this
constraint, it has emerged that the most favorable configuration has a 
stellar--mass black hole ($M_{BH} \approx 3-10 M_\odot$) surrounded by a thick
torus of matter ($M_t \approx 0.01-1 M_\odot$, with $\rho \approx 10^{10}\; g
\; cm^{-3}$). The presence of a black hole is {\it not} required by observations
in any way: models involving neutron stars are still admissible, the advantage 
of having a black hole being only the deeper potential well: you get more 
energy out per unit accreted mass. The configuration thusly envisaged has 
a cone surrounding the symmetry axis devoid of baryons, since all models
leading to this configuration have large amounts of specific angular momentum,
and thus baryons close to the rotation axis either are not there, or have
accreted onto the black hole due to their lack of centrifugal support. 

\subsection{Energy release mechanism}

There are two major mechanisms for energy release discussed in the literature,
the first one to be proposed (Berezinsky and Prilutskii 1986) being the reaction
$\nu +\bar{\nu}\rightarrow e^-+e^+$. Neutrinos have non--negligible mean free
paths in the tori envisaged here, so that this annihilation reaction will 
take place not inside tori themselves, where they are preferentially generated
because densities are highest, but in a larger volume surrounding the source. 
This is both a blessing and a disgrace: by occupying a larger volume, the 
probability that every neutrino finds its antiparticle to annihilate decreases,
but then the energy is released in baryon--cleaner environments. The problem,
though complex, is eminently suitable for numerical simulations, showing
(Janka \etal, 1999, and references therein) that about $10^{50}\; ergs$ can be 
released this way, above the poles of a black hole where less than $10^{-5} 
M_\odot$ are found. 

Highly energetic bursts cannot be reproduced by this mechanism, due to its
low efficiency: the second mechanism proposed involves the conversion of
Poynting flux into a magnetized wind. The basic physical mechanisms are 
well--known (Usov 1992) since they have been studied in the context of 
pulsar emission: electrons are accelerated by a motional electric field
$\vec{E} = \vec{v}\wedge\vec{B}/c$ due to the rotation of a sufficiently 
strong magnetic dipole, attached either to a black hole, or to the torus. 
Photons are then produced by synchrotron or curvature
radiation, and photon/photon collisions produce pairs, to close the circle and
allow looping. In order to carry away $10^{51}\; erg\; s^{-1}$, a magnetic
field of $\approx 10^{15}\; G$ is required. This is not excessive, since 
it is about three orders of magnitude below equipartition with torus matter, 
and because such fields already exist in nature, see SGR 1806-20 and SGR
1900+140: the key point is to understand whether some kind of dynamo effect
can lead to these high values within the short allotted time. 

Depending upon whether the open magnetic field lines extending to infinity
are connected to the black hole or to the torus, the source of the energy
of the outflow will be the rotational energy of the black hole (the so--called
Blandford--Znajek effect) or of the torus. The first case is traditionally
discussed in the context of AGNs (Rees, Blandford, Begelman and Phinney 1984),
but it is harshly disputed whether the energy outflow may be actually 
dominated by the black hole rather than by the disk (Ghosh and Abramowicz, 
1997,  Livio, Ogilvie and Pringle 1998). On the other hand, the torus looks 
ideal as the source
of a dynamo: its large shear rate, the presence of the Balbus-Hawley 
instability to convert polidal into toroidal flux, and the possible presence
of the anti--floating mechanism inhibiting ballooning of the magnetic field
(Kluzniak and Ruderman 1998), all seem to favor the existence of a fast 
dynamo. It should also be remarked that the configuration of the magnetic
field in this problem is known: in fact, the configuration discussed in 
Thorne \etal, 1986 for black holes, only uses the assumptions of 
steady--state and axial symmetry, and is thus immediately extended to magnetic
fields anchored to the torus. What is really required here is a first order 
study, of the sort published by Tout and Pringle (1992) on angular momentum 
removal from young, pre--main--sequence stars via magnetic stresses, and on the 
associated $\alpha-\omega$ dynamo. Until such studies are made, it will be 
premature to claim that neutrino annihilations are responsible for the powering 
of GRBs. 

\subsection{Progenitors}

There is no lack of proposed progenitors, but I will discuss only 
binary neutron mergers (Narayan, Paczy\`nski and Piran 1992), collapsars
(Woosley 1993, Paczy\`nski 1998) and SupraNovae (Vietri and Stella 1998, 1999). 

Clearly, NS/NS mergers is the best model on paper: it involves objects 
which have been detected already, orbital decay induced by gravitational
wave emission is shown by observations to work as per the theory, and numerical
simulations by Janka's group show that a neutrino--powered outflow 
in baryon--poor matter can be initiated. The major theoretical uncertainties
here concern bursts' durations and energetics: all numerical models produce
short bursts ($\approx 0.1\; s$) with modest energetics, $E < 10^{51}\; erg$.
This is a direct consequence of the mechanism for powering the burst: large, 
super--Eddington luminosities are carried away by neutrinos, leading to a
large mass influx, but only a small fraction, $1-3\%$, can be harnessed for the
production of the burst. Furthermore, we cannot invoke large beaming factors in 
this case: the outflow is only marginally collimated, in agreement with 
expectations that an accretion disk with inner and outer radii $R_{out}/R_{in} 
\approx $ a few (for the case at hand) can only produce a beam semi--opening 
angle of $R_{in}/R_{out}$. So, perhaps, this model may account for the short
bursts, but it should be remembered that nothing of what was discussed above
pertains to this subclass: BeppoSAX (and thus all BeppoSAX--triggered 
observations) can only detect long bursts. 

On the other hand, future space missions, whether or not able to locate short 
bursts, can provide a decisive test of this model, provided they can follow
with sufficient sensitivity a given burst for several hours. This model, in 
fact, is the only one proposed so far according to which some explosions should 
take place outside galaxies: according to Bloom, Sigurdsson and Pols (1999), 
about $50\%$ of all bursts will be located more than $8\; kpc$ from a galaxy, 
and $15\%$ in the IGM. This characteristic is testable without recourse to 
optical observations. In fact, the afterglow begins with a delay (as seen by an 
outside observer) of $t_d = (r_{ag}-r_{sh})/\gamma^2 c \approx r_{ag}/\gamma^2 
c$, which varies greatly depending upon the environment in which the burst 
takes place:
\begin{equation}
t_d = \left\{ 
\begin{array}{ll}
15 \; s & \mbox{ISM, n =}1 \; cm^{-3} \\
5 \; min & \mbox{galactic halo, n = }10^{-4}\; cm^{-3} \\
4\; h & \mbox{IGM, n = }10^{-8}\; cm^{-3}
\end{array}
\right.
\end{equation}
Between the burst proper and the beginning of the power--law--like afterglow,
thus a silence of recognizable duration is expected (Vietri 2000b). 

Collapsars are currently in great vogue as a possible source of GRBs: the 
large amoung of energy available as the core of a supermassive star collapses 
directly to a black hole is in fact very attractive, even though (again!) the
limited efficiency of the reaction $\nu+\bar{\nu}\rightarrow e^- + e^+$ makes
most of this energy unavailable. Here too
there is some evidence that these objects must exist (Paczy\`nski 1998), and
numerical simulations again showing energy preferentially deposited along 
the hole rotation axis are also available (McFayden and Woosley 1999). Here
however, what is truly puzzling is how the outflow can pierce the star's outer 
layers without loading itself with baryons: we should remember that at most
$10^{-6} M_\odot$ can be added to $10^{51}\; erg$: more baryons imply a 
proportionately slower outflow. The argument is that the dynamical timescale
of the outer layers of a massive stars is of order of a few hours, so that,
even if the core collapses and pressure support is removed, nothing will 
happen during the energy release phase: the outflow must pierce its way 
through. Two processes seem especially dangerous: Rayleigh--Taylor instability
of the fluid heated--up by neutrino annihilations as it is weighed upon by the
colder, denser outer layers, and Kelvin--Helmholtz instability after the hot 
fluid has pierced the outer layers and is passing through the hole. It is 
well--known that the non--linear development of these instabilities leads to
mass entrainment, and that the time--scale for the development of these 
instabilities is very fast. Furthermore, the baryon--free outflow may be 
`poisoned' by baryons to a deadly extent, even if numerical simulations, with 
their finite resolution, were to detect nothing of the kind. 

The third class of models, SupraNovae, concerns supramassive neutron stars which
are stabilized against self--gravity by fast rotation, to such an extent 
that they cannot be spun down to $\omega = 0$ because they implode to a 
black hole. As the star's residual magnetic dipole sheds angular momentum, this
is exactly the fate to be expected for the whole star, except for a small
equatorial belt , whose later accretion will power the burst. It is easy to
show that this implosion must take place in a very baryon--clean environment.
The major uncertainties here concern the channels of formation and the existence
of this equatorial belt. Two channels of formation have been proposed: 
direct collapse to a supramassive configuration (Vietri and Stella 1998) and 
slow mass accretion in a low--mass X--ray binary (Vietri and Stella 1999). 
Both are possible, though none yet is supported by observations. The existence
of the left--over belt has recently been questioned by Shibata, Baumgarte and
Shapiro (1999), who however simulated the collapse of neutron stars with 
intermediate equations of state, which are entirely (or nearly exactly so)
contained inside the marginally stable orbit even before collapse: clearly,
these must be swallowed whole by the resulting black hole. Soft equations of
state are free of this objection, and are thus much more likely to leave behind
an equatorial belt. The soft EoSs are especially favored since the neutron 
stars must survive the $r$--mode instability, and thus soft EoSs (Weber 
1999) would be in any case required. So one might say that the existence of 
these stars hinges on one uncertainty only, the EoS of nuclear matter. 
Besides the baryon--clean environments, SupraNovae have another advantage 
over rival modlels: only the lowest density regions would be left behind, 
precisely those with the smallest neutrino losses. The powering of the burst
can thus occur through accretion caused by removal of angular momentum by 
magnetic stresses, without the parallel, unproductive, neutrino generation.


\section{Conclusions}

It is difficult to end on an upbeat note: we cannot expect in the near future 
a rate of progress similar to the one we witnessed in the past three years. 
In particular, it may be expected that the next flurry of excitement will
come with the beginning of the SWIFT mission, which promises to collect
relevant data (redshifts, galaxy types, location within or without galaxies,
absorption or emission features in the optical and in the X--ray) for a few
hundred bursts. This data will nail the major characteristics of the environment
(at large) in which bursts take place, and we may be able to rule out a few 
models. On the other hand, the energy release process, shrouded as it is
in optical depths $> 10^{10}$, will remain mysterious, our only hope in this
direction being gravitational waves. 

Judging by the analogy with radio pulsars, this will correspond to the 
flattening of the learning curve. Aside from this, we may hope to locate the 
equivalent of the binary radio pulsar, but, differently from Jo Taylor, we 
have to be awfully quick in grabbing it.

\begin{acknowledgements}
Thanks are due to Gabriele Ghisellini, who wisely steered me away from 
synchrotron emission, and toward the true light of Inverse Compton. 
\end{acknowledgements}

\end{document}